\documentclass[procedia]{easychair}

\usepackage{doc}
\usepackage{makeidx}

\usepackage{epstopdf}
\hypersetup{%
    pdfborder = {0 0 0}
}

\def\procediaConference{INTERNATIONAL CONFERENCE ON COMPUTATIONAL SCIENCE (ICCS 2015)}

\title{Coevolution of Information Processing and Topology 
in Hierarchical Adaptive Random Boolean Networks}

\titlerunning{Coevolution of Information Processing and Topology}

\author{
    Piotr J. G\'{o}rski\inst{1}
\and
    Agnieszka Czaplicka\inst{1}
\and
    Janusz A. Ho{\l}yst\inst{1,2}
}

\institute{
  Faculty of Physics, Center of Excellence for Complex Systems Research, Warsaw University of Technology, Koszykowa 75, PL-00-662 Warsaw, Poland
  Warsaw, Poland\\
  \email{piogor@student.if.pw.edu.pl}, \email{jholyst@if.pw.edu.pl}
\and
   ITMO University,
   19, Kronverkskiy av., 197101 Saint Petersburg, Russia\\
 }

\authorrunning{G\'{o}rski, Czaplicka and Ho{\l}yst}

\begin{document}

\maketitle


\begin{abstract}
Random Boolean networks (RBNs) are frequently  employed for modelling  complex systems driven by information processing, e.g. for gene regulatory networks (GRNs). Here we propose a hierarchical adaptive RBN (HARBN) as a system consisting of distinct adaptive RBNs –-- subnetworks –-- connected by a set of permanent interlinks. Information measures and internal subnetworks topology of HARBN coevolve  and reach steady-states that  are specific for a given network structure. We investigate mean node information, mean edge information as well as a mean node degree as functions of model parameters and demonstrate HARBN’s ability to describe complex hierarchical systems. 
\end{abstract}


%
%

\section{Introduction}
\label{sect:introduction}

Complex networks are frequently used to describe evolving, constantly changing objects that consist of elements and complicated functional relations between them. Simplified models were created to study and compare certain network properties. One of the examples are random Boolean networks (RBNs). RBNs are generic \cite{gershenson04} and that is why they have been applied in many different fields. 

Gene regulatory networks (GRNs) are models describing the structure and behavior of the transcriptional network responsible for regulating the gene expression in a cell \cite{costa08, mnw}. GRNs are both robust and stochastic. GRNs  must respond to diverse stimuli and they  are highly modular, e.g.  hierarchical regulatory interactions in transcriptional networks in yeast. Analysis of GRNs provided evidence that diverse stimuli may cause modifications of gene interactions and network topology. During GRN evolution edges are created and destroyed, and  in this way the gene expression is altered. 

In \cite{mlb} Liu and Bassler proposed the RBN modification, where  network topology is time dependent and the observed coevolution encompasses some effects noticed in GRN. Here, we extend the original Liu and Bassler's model, most of all by incorporating the concept of hierarchy. In our approach  there are no  robust genes, but a part of network edges are assumed to be robust, i.e. they stay unchanged during the system evolution.

The paper is organised as follows: in section \ref{sect:model} we describe our model, i.a. we present  the algorithm of network evolution (section \ref{sect:algorithm}). Section \ref{sect:information} presents the applied information measures. Section \ref{sect:simulations} contains simulations description, results and discussion. Finally, in section \ref{sect:conclusions} we conclude our studies.

\section{Model of Hierarchical Adaptive Random Boolean Network (HARBN)}
\label{sect:model}
Let us consider a system of coupled nodes in a form of a directed network that corresponds to a GRN. For simplicity we will assume that internal nodes' variables $\sigma_n(t)$ are $0$ or $1$ and they can evolve in time according to randomly chosen boolean functions $f_i$. Arguments of the function $f_n$ are internal variables $\sigma_m(t)$  of all nodes $m$ such that there is a connection from a  node $m$ to the node $n$.

During simulations of adaptive RBN (ARBN) changes of network state and topology are measured. The simulation consists of defined {\it a priori} number of epochs. In each epoch the network's attractor is found (i.e a periodic orbit of variables $\sigma_n(t)$ that is reached by the system after a transient time), one node is randomly chosen and according to the Activity-dependent Rewiring Rule (ADRR) (described in section \ref{sect:rule}) a number of incoming connections to this node is changed. 

We extended Liu and Bassler's model. First, we consider a small number of ARBNs, called subnetworks. Different subnetworks are sparsely connected by directed edges, called interlinks. We build our model in such a way that the considered number of interlinks is smaller than the number of edges within the subnetworks. Therefore, the created network is hierarchical in terms of connection density, because certain groups of nodes are much denser connected in internal blocks (subnetworks)  as compared to  the rest of the network. During a single epoch the network's attractor is found and for each subnetwork a node is chosen randomly. Connections of such a node are changed according to the ADRR. The edges may arise only between nodes in the same subnetwork. The interlinks are permanent edges: they are created at the beginning of each simulation and they cannot  be changed. Secondly, we also alter the ADRR by introducing a resilience parameter $\alpha$ (see section \ref{sect:rule}). In the original ADRR the 
resilience parameter did not exist and it corresponds to $0$ in our approach.

\subsection{Activity-Dependent Rewiring Rule}
\label{sect:rule}

Let us define the network resilience parameter that will decide on the type of  changes in network  topology as $\alpha$ where $\alpha \in [0,0.5]$.

If the node's mean state $\left\langle \sigma_n\right\rangle$  during the attractor is not higher than $\alpha$ or at least $(1-\alpha)$, then this node is considered to be frozen and one new incoming edge is added to this node. The new edge starts in a random node chosen from the group of nodes from the same subnetwork that does not already have a link to the considered node. Otherwise, this node is considered to be active and one of its edges is randomly chosen and deleted.

For non-zero values of parameter $\alpha$ the original ADRR is extended so as nodes with a mean state higher than $0$ or smaller than $1$ are considered to be frozen.

\subsection{Adaptive Algorithm for Hierarchical RBNs}
\label{sect:algorithm}
Our algorithm for the coevolution of hierarchical RBNs is as follows:
\begin{enumerate} 
\item Generate $M$ uniform Boolean networks (subnetworks) containing $N_M$ nodes each with $K_i$ directed incoming edges starting in a randomly chosen group of nodes belonging to the same subnetwork. Then generate $K_M$ edges between subnetworks (interlinks). (For each node generate necessary Boolean functions with a bias parameter $p=1/2$.)
\item Generate a random initial state $S(0)$ of all internal node variables $\sigma_n(0)$  and find the network's attractor length using  the following algorithm \cite{mlb}:
\begin{description}
\item [$a.$] Define comparison moments. Here: \\ $\boldsymbol{T}=\{0; 100; 1,000; 10,000; 100,000\}$. Set $k=0$ and $i=1$.
\item [$b.$] Synchronically update the states of the network's nodes: $S(i)$. If $S(i)$ equals $S(T_k)$ the attractor length is $(i - T_k)$. Go to point 3. Otherwise continue.
\item [$c.$] If $i$ equals $T_{k+1}$, increment $k$. If $k$ is higher than $4$ end this subalgorithm with the attractor length equal to $(T_4 - T_3)$. Otherwise, increment $i$ and go back to point 2b. 
\end{description}
\item Choose a node from each subnetwork and calculate its mean state during one attractor cycle. 
\item According to the ADRR, change the topology of each subnetwork. Do not delete interlinks. In case there are no connections to delete, randomly choose another node and repeat this point. 
\item Generate new Boolean functions for each node.
\item If the predefined number of epochs is not reached, go back to point $2$. 
\end{enumerate}
Remarks:
\begin{itemize}
\item The interlinks are generated as follows: first the highest possible number of interlinks is equally distributed between ordered pairs  of subnetworks (due to directed edges each pair of subnetworks appear twice); secondly, the remaining interlinks are generated between random pairs. 
\item The described method allows finding the attractor length equal to $90,000$ at most.
\item Advancing through the steps 2-6 is called an epoch (after \cite{mlb}). Updating the states of all nodes in step 2b. is called an iteration. 
\end{itemize}

\section{Information Measures of RBNs}
\label{sect:information}

Measuring information amount transmitted by a network is an important tool that  facilitates exploring system's features. There are many approaches to do it, see e.g. \cite{blf,pks}. Here we define network {\it activity information} $I$ as a sum of transformed activities of each node. Activity $A_n$ of a node $n$ is a number of changes of the node's state during the network attractor divided by the attractor length. There is no information stored in frozen nodes (nodes with $A_n=0$) and in nodes whose state changes in every iteration ($A_n=1$). Such nodes should not contribute to the network information. On the other hand, nodes which change their state the same number of times as they keep the old state ($A_n=0.5$) contribute mostly because of their unexpected behavior. Node activity can be identified with probability of changing the node's state. Then the activity $A_n$ links with the parameter $a$ described in \cite{blf} as system's "self-overlap": $A_n=1-a$. Here we introduce a natural definition of node 
{\it activity information} as: 
\begin{equation}
I_n=-A_n\log(A_n)-(1-A_n)\log(1-A_n)
\end{equation}
The total network activity information is given as $I=\sum_{n} I_n$. For brevity  we further write {\it network information} instead of network activity information. 
\section{Simulations}
\label{sect:simulations}

\begin{figure}[tb]
	\begin{centering}
	\includegraphics[width=0.5\textwidth]{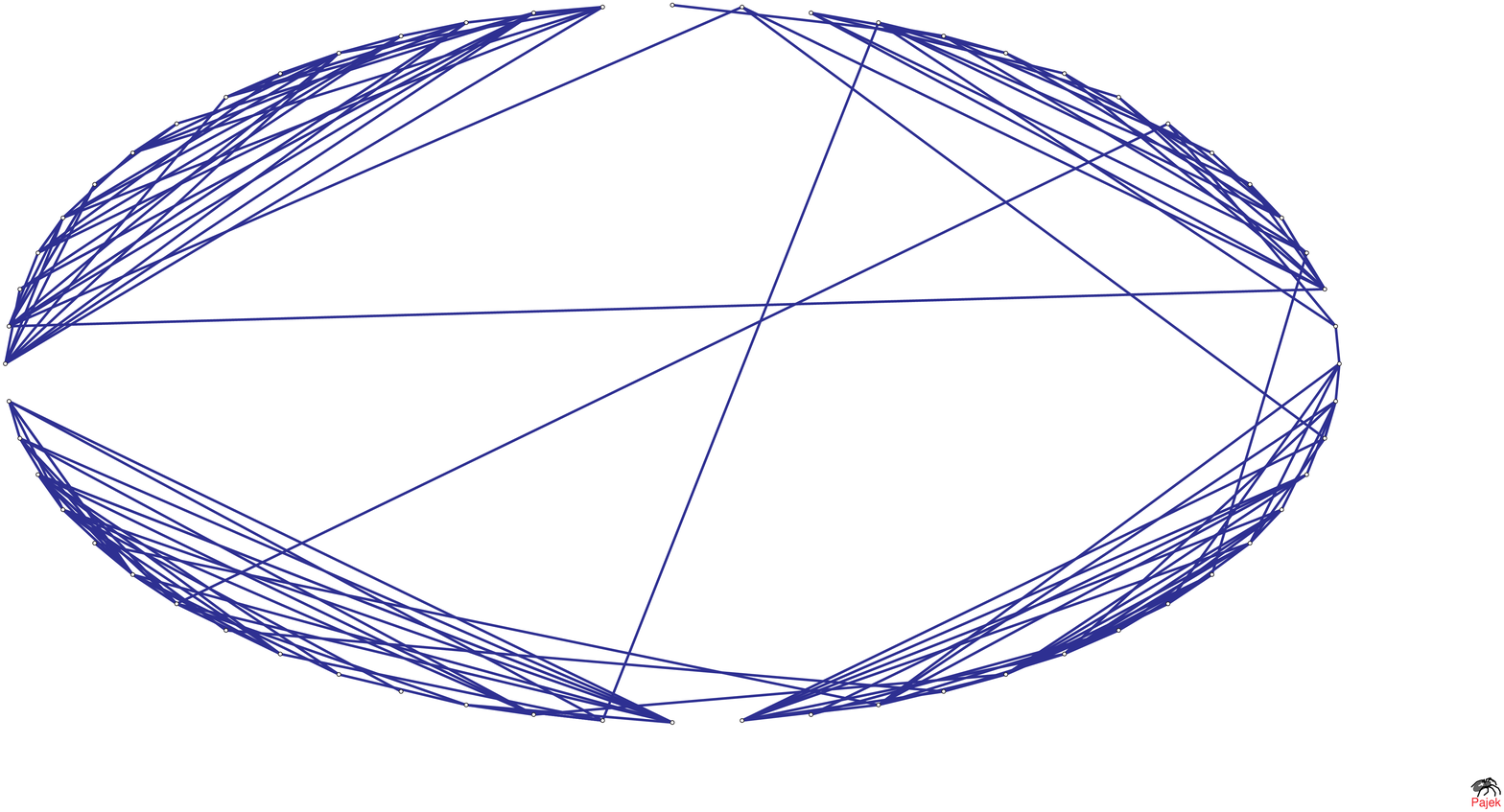}
	\caption{Example structure after $1000$ epochs of a network $4x15x10$.}
	\label{fig:example}
	\end{centering}
\end{figure}

The simulations of different structures of HARBN model consisted of $1000$ epochs repeated $100$ times. We let $MxN_M xK_M$ denote a network consisting of $M$ subnetworks each of $N_M$ nodes and linked by $K_M$ interlinks. Here we show the results for $60-$ and $80-$node networks. Our simulations were conducted in $3$ parts. In the beginning we reproduced the results achieved in \cite{mlb} ($M=1$, $\alpha=0$). Then, we divided networks into $2-4$ subnetworks and linked them by $0-80$ interlinks. We compared the results for ARBNs and HARBNs. At the end the system properties for non-zero resilience values $\alpha$ were explored. An example structure of simulated network after $1000$ epochs is shown in Figure \ref{fig:example}.

Each realization of ARBNs and HARBNs demonstrates different  structure and information parameters. However identical networks tend to oscillate (after the initial transient period) near the same mean steady-state (m.s.s.) levels. On the other hand, networks of different number of nodes, subnetworks or interlinks tend to reach different m.s.s. values. Therefore, for each network type we define: m.s.s. incoming connectivity $K_{ss}$, m.s.s. network information $I$, m.s.s. node information $IPN$ (calculated  as information $I$ per node), m.s.s. edge information $IPE$ (calculated  as information $I$ per edge) and geometric m.s.s. attractor length $T$. If not other mentioned we assume an arithmetic mean. In order to determine all above parameters $200$ beginning epochs of each realization were discarded as transient periods. $K_{ss}$ is calculated as follows: 
\begin{itemize}
\item calculate mean incoming connectivity in an epoch $\rightarrow$ $K_{in}$, 
\item calculate mean $K_{in}$ starting from 201th epoch in a realization $\rightarrow$ $\left<K_{in}\right>$, 
\item calculate mean $\left<K_{in}\right>$ over all realizations $\rightarrow$ $K_{ss}$. 
\end{itemize}
The information parameters $I$ and $IPN$ are calculated likewise. $IPE$ is computed by dividing $IPN$ by $K_{ss}$. 

Now we shall discuss the influence of network topology on system properties. We are aware of the fact that  because of  a small number of investigated  networks additional large-scale simulations will be necessary to confirm our findings. 

\subsection{Mean Steady-State Connectivity}
\label{sect:mss-con}

\begin{figure}[tb]
	\begin{centering}
	\includegraphics[width=0.9\textwidth]{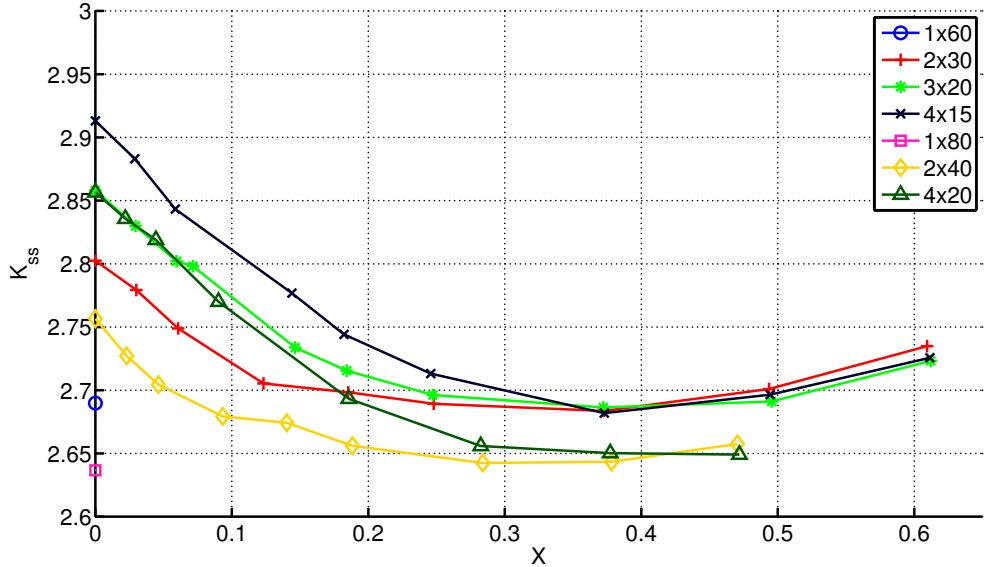}
	\caption{The mean steady-state connectivity $K_{ss}$ as a function of the ratio of the interlinks to the total number of  edges for various networks.}
	\label{fig:Kss-X}
	\end{centering}
\end{figure}

Let $X$ denote the ratio of the interlinks to the total number of  edges. For small $X$ values different network types differentiate themselves in terms of $K_{ss}$  by the size of a subnetwork (Figure \ref{fig:Kss-X}) and $K_{ss}$ decreases with the growth of the subnetwork size. With the increase of $X$ the subnetworks become more and more linked. For high $X$ values different network types start to differentiate themselves by the size of the whole network, i.e. subnetworks are strongly mutually dependent. We estimate that by $X\approx0.40$ distinct subnetworks do not exist anymore. The resulting $K_{ss}$ values are independent from the subnetwork size. 

\subsection{Information per Node}
\label{sect:ipn}

\begin{figure}[tb]
	\begin{centering}
	\includegraphics[width=0.9\textwidth]{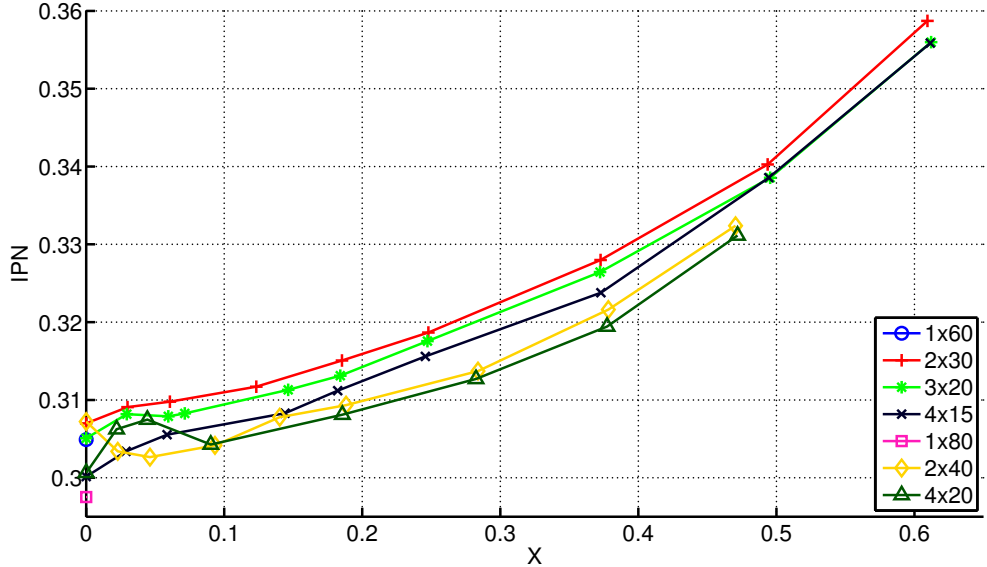}
	\caption{The mean node information $IPN$ as a function of  the ratio of the interlinks to the total number of edges for various networks.}
	\label{fig:IPN-X}
	\end{centering}
\end{figure}

Calculated information per node $IPN$ also exhibits different ordering for small and large $X$ values (Figure \ref{fig:IPN-X}). For $X \ll 0.1$ networks (apart from those with one subnetwork) differentiate with the increase of the number of distinct subnetworks. When  $X >0.1$  then the information $IPN$ differentiates the networks by the size of the whole network. $IPN$ is larger in smaller networks. In the interval  $0<X<0.1$, especially for $80-$node networks, ambiguous behaviour is observable.  For the $4x20$ network the information $IPN$ starts at a rather low value. A small increase of the ratio $X$ causes a quick rise of $IPN$ with a maximum for $X\approx 0.05$. Then until $X\approx 0.1$ a moderate decrease of $IPN$  is observable. The behaviour of $IPN$ for $2x40$ is different. $IPN$ starts  from a high value for $X=0$  and then quickly falls to a minimum in $X\approx 0.05$. Afterwards a stable increase is visible. 

\subsection{Information per Edge}
\label{sect:ipe}

\begin{figure}[tb]
	\begin{centering}
	\includegraphics[width=0.9\textwidth]{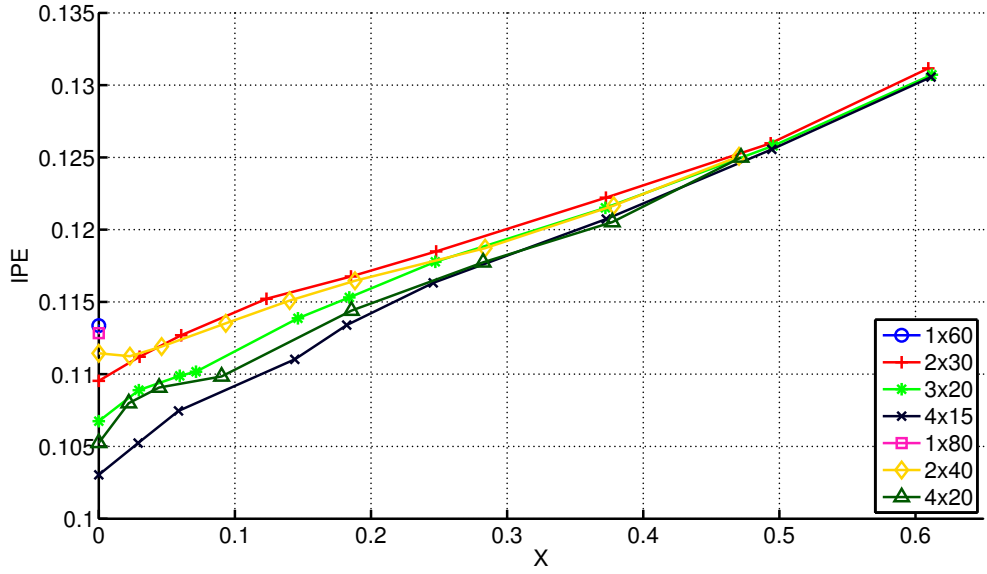}
	\caption{The mean edge information $IPE$ as a function of the  ratio of the interlinks to the total number of  edges  for various networks. $IPE$ has been calculated by dividing the mean node information $IPN$ by the mean steady-state in-degree connectivity $K_{ss}$.}
	\label{fig:IPE-X}
	\end{centering}
\end{figure}

The behaviour of the third network parameter --- $IPE$ (Figure \ref{fig:IPE-X}) includes some features of the previous two observables. For $X\approx0$ networks differentiate themselves first by the number of distinct subnetworks (less subnetworks, higher $IPE$), secondly by the size of an  individual subnetwork (with one exception for networks $1xN$). On the other hand, ordering by the network size for higher $X$ is very weak and it appears that $IPE$ is equal for all systems investigated by the same $X$ value. It does not depend on the total network size or the number of subnetworks.

\subsection{Geometric Mean Attractor Length}
\label{sect:references}

\begin{figure}[ht]
	\begin{centering}
	\includegraphics[width=0.9\textwidth]{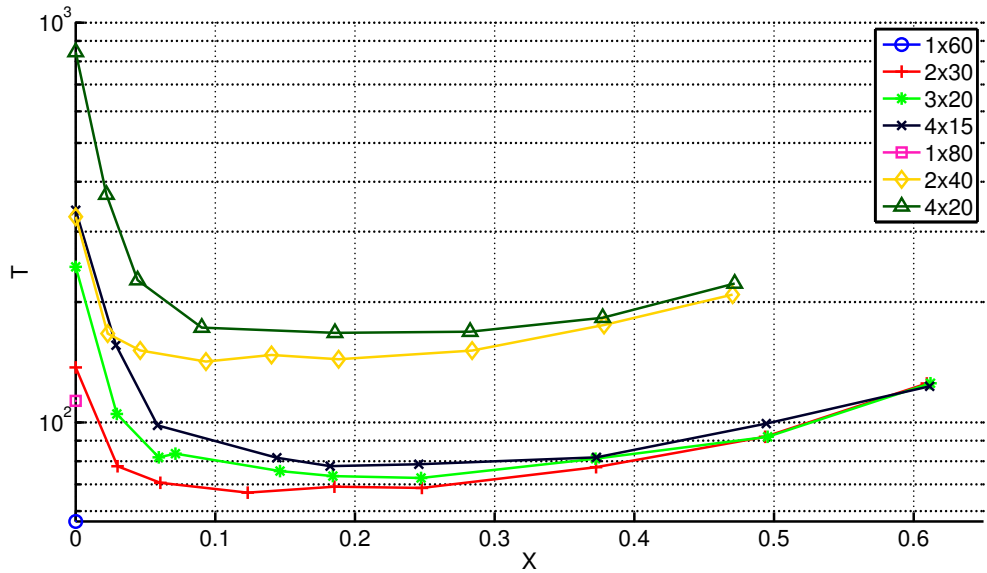}
	\caption{The geometric mean attractor length $T$ as a function of the ratio of the interlinks to the total number of edges for various networks.}
	\label{fig:T-X}
	\end{centering}
\end{figure}

Figure \ref{fig:T-X} shows the relationship between the geometric mean attractor length $T$ and the ratio $X$. The lengths of the attractors grow with the increase of the total number of nodes. Moreover, structures of the same size with more subnetworks tend to possess longer attractors. Dividing a network into subnetworks creates separate unconnected parts ($X \approx 0$). Such structures reach the highest $T$ values. It is related to the attractor search subalgorithm: the attractor is found for the whole network and its separate parts multiply and elongate the attractor length. Interlinks connect the subnetworks, but interlinks are not flexible. Interlinks significantly decrease the attractor length leading to higher flexibility. The smallest attractors are achieved for $X$ ranging from around $0.1$ to around $0.25$. Further increase of the interlinks ratio leads to longer lengths $T$, as interlinks stiffen the network.

\subsection{Resilience Parameter}
\label{sect:resilience}

\begin{figure}[ht]
	\begin{centering}
	\includegraphics[width=0.9\textwidth]{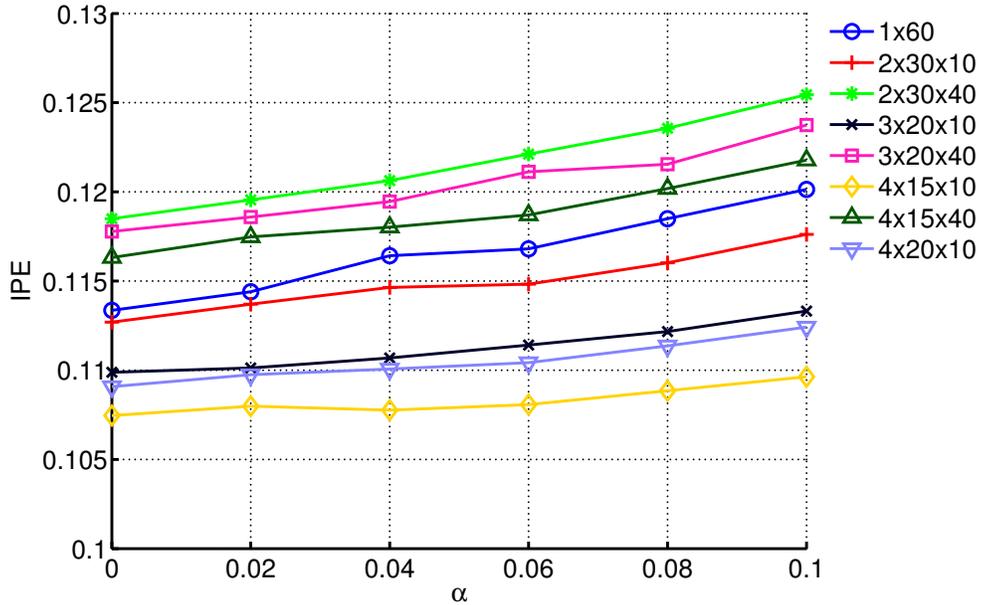}
	\caption{The mean edge information $IPE$ as a function of the resilience parameter $\alpha$ for several different network types. $IPE$ has been calculated by dividing the mean node information $IPN$ by the mean steady-state in-degree connectivity $K_{ss}$.}
	\label{fig:IPE-a}
	\end{centering}
\end{figure}

In order to explore non-zero values of the resilience parameter $\alpha$ different network structures with $10$ and $40$ interlinks were analysed. These numbers of interlinks  corresponded to $X\approx 0.05$ and $X\approx 0.25$ respectively. Effects of hierarchical structures and interactions between different subnetworks were observed. Figure \ref{fig:IPE-a} shows that the parameter $\alpha$ differentiates networks of different types. First of all we can distinguish number of interlinks: there are $3$ levels for $0$, $10$, and $40$ interlinks. Their order is primarily the result of values demonstrated in Figure \ref{fig:IPE-X}. Within the networks of the same number of interlinks higher mean edge information corresponds to networks with larger subnetworks. Comparing $3x20x10$ and $4x20x10$ networks we can see the next level of ordering: higher $IPE$ is in a smaller network. We have also observed (it is not shown in the paper) that  non-zero values of $\alpha$ lead to higher $K_{ss}$ values. Therefore the 
information per node $IPN$ grows with the steady-state-connectivity $K_{ss}$  faster than the information per edge $IPE$.

\section{Conclusions}
\label{sect:conclusions}

The model of hierarchical adaptive random Boolean network (HARBN) has been introduced and numerically explored. The system consists of subnetworks connected by fixed interlinks, where the internal topology of the subnetworks can evolve depending on individual node activity. We have observed that the main natural feature of ARBNs, i.e. their adaptability is preserved in HARBNs that can evolve towards stable configurations. When the ratio $X$ of interlinks to the total number of edges is of the order of $X\approx 0.4 $ then the interlinks efficiently re-connect the whole network (as if separate parts did not exist). However such a large system is less flexible and displays longer attractors. The shortest attractors are observable for $X$ ranging from $0.1$ to $0.25$ depending on the network structure.

The mean node information ($IPN$) and the mean edge information ($IPE$) grow in HARBN with the increase of the ratio $X$ as well as with the resilience parameter $\alpha$ and $IPE$ tends to achieve the same value for all network types in case of many interlinks. Adding a new node to the network decreases $IPN$ and, moreover, leads to fewer incoming connections per node. On the other hand adding a new interlink increases $IPE$ values. We conclude that the introduced HARBNs may successfully be used to model GRNs with a modular structure and they well describe the processes of evolution and speciation. 

\subsection{Acknowledgements}

The research leading to these results has received funding from the European Union Seventh Framework Programme (FP7/2007-2013) under Grant Agreement No. 317534 (the Sophocles project) and from the Polish Ministry of Science and Higher Education Grant No. 2746/7.PR/2013/2. J.A.H. has been also partially supported by Russian Scientific Foundation, proposal $\#$14--21--00137 and by European Union COST TD1210 KNOWeSCAPE action.

%
\label{sect:bib}
\bibliographystyle{unsrt}
\bibliography{rbn-iccs3}

\begin{thebibliography}{1}

\bibitem{gershenson04}
C.~Gershenson.
\newblock Introduction to random boolean networks.
\newblock In {\em 9th International Conference on the Simulation and Synthesis
  of Living Systems}, pages 160--173, 2004.

\bibitem{costa08}
F.A.~Rodrigues L.~da F.~Costa and A.S. Cristino.
\newblock Complex networks: The key to systems biology.
\newblock {\em Genetics and Molecular Biology}, 31(3):591--601, 2008.

\bibitem{mnw}
L.T. MacNeil and A.J.M. Walhout.
\newblock Gene regulatory networks and the role of robustness and stochasticity
  in the control of gene expression.
\newblock {\em Genome Research}, 21(645), 2011.

\bibitem{mlb}
M.~Liu and K.E. Bassler.
\newblock Emergent criticality from coevolution in random boolean networks.
\newblock {\em Physical Review E}, 74(041910), 2006.

\bibitem{blf}
B.~Luque and A.~Ferrera.
\newblock Measuring mutual information in random boolean networks.
\newblock {\em Complex Systems}, 12:241--252, 2000.

\bibitem{pks}
P.~Krawitz and I.~Shmulevich.
\newblock Basin entropy in boolean network ensembles.
\newblock {\em Physical Review Letters}, 98(158701), 2007.

\end{thebibliography}


\end{document}